\documentclass{aastex63}

\received{October 1, 2021}
\revised{December 3, 2021}
\accepted{December 12, 2021}
\submitjournal{PSJ}
\usepackage{lineno}
\shorttitle{Col-OSSOS}
\shortauthors{Buchanan et al.}
\graphicspath{{./}{figures/}}

\newcommand\revision[1]{#1}

\begin{document}

\title{Col-OSSOS: Probing Ice Line/\revision{Color} Transitions within the Kuiper Belt's Progenitor Populations}

\correspondingauthor{Laura E. Buchanan}
\email{lbuchanan14@qub.ac.uk}

\author[0000-0002-8032-4528]{Laura E. Buchanan}
\affiliation{Astrophysics Research Centre, Queen's University Belfast}

\author[0000-0003-4365-1455]{Megan E. Schwamb}
\affiliation{Astrophysics Research Centre, Queen's University Belfast}

\author[0000-0001-6680-6558]{Wesley C. Fraser}
\affiliation{NRC-Herzberg Astronomy and Astrophysics, National Research Council of Canada, 5071 West Saanich Road, Victoria, BC V9E 2E7, Canada}

\author[0000-0003-3257-4490]{ Michele T. Bannister}
\affiliation{School of Physical and Chemical Sciences—Te Kura Matū, University of Canterbury, Private Bag 4800, Christchurch 8140, New Zealand}

\author[0000-0001-8617-2425]{Michaël Marsset}
\affiliation{Department of Earth, Atmospheric and Planetary Sciences, MIT, 77 Massachusetts Avenue, Cambridge, MA 02139, USA}

\author[0000-0003-4797-5262]{Rosemary~E. Pike}
\affiliation{Center for Astrophysics $|$ Harvard \& Smithsonian, 60 Garden Street, Cambridge, MA 02138, USA}

\author[0000-0002-4547-4301]{David Nesvorný}
\affiliation{Department of Space Studies, Southwest Research Institute, 1050 Walnut St., Suite 300, Boulder, CO, 80302, United States}

\author[0000-0001-7032-5255]{J. J. Kavelaars}
\affiliation{NRC-Herzberg Astronomy and Astrophysics, National Research Council of Canada, 5071 West Saanich Road, Victoria, BC V9E 2E7, Canada}
\affiliation{Department of Physics and Astronomy, University of Victoria, Elliott Building, 3800 Finnerty Road, Victoria, BC V8P 5C2, Canada}

\author[0000-0001-8821-5927]{Susan D. Benecchi}
\affiliation{Planetary Science Institute, 1700 East Fort Lowell, Suite 106, Tucson, AZ 85719, USA}

\author[0000-0003-4077-0985]{Matthew J. Lehner}
\affiliation{Institute of Astronomy and Astrophysics, Academia Sinica; 11F of AS/NTU Astronomy-Mathematics Building, No.1, Sec. 4, Roosevelt Rd Taipei 10617, Taiwan, R.O.C}

\author{Shiang-Yu Wang}
\affiliation{Institute of Astronomy and Astrophysics, Academia Sinica; 11F of AS/NTU Astronomy-Mathematics Building, No.1, Sec. 4, Roosevelt Rd Taipei 10617, Taiwan, R.O.C}

\author[0000-0002-6830-476X]{Nuno Peixinho}
\affiliation{Instituto de Astrofísica e Ciências do Espaço, Universidade de Coimbra, 3040-004 Coimbra, Portugal}

\author[0000-0001-8736-236X]{Kathryn Volk} 
\affiliation{Lunar and Planetary Laboratory, University of Arizona, 1629 E. University Blvd., Tucson, AZ 85721, USA}

\author[0000-0003-4143-8589]{Mike Alexandersen}
\affiliation{Center for Astrophysics $|$ Harvard \& Smithsonian, 60 Garden Street, Cambridge, MA 02138, USA}

\author[0000-0001-7244-6069]{Ying-Tung Chen}
\affiliation{Institute of Astronomy and Astrophysics, Academia Sinica; 11F of AS/NTU Astronomy-Mathematics Building, No.1, Sec. 4, Roosevelt Rd Taipei 10617, Taiwan, R.O.C}

\author{Brett Gladman}
\affiliation{Department of Physics and Astronomy, University of British Columbia, Vancouver, BC, Canada}

\author{Stephen Gwyn}
\affiliation{NRC-Herzberg Astronomy and Astrophysics, National Research Council of Canada, 5071 West Saanich Road, Victoria, BC V9E 2E7, Canada}

\author[0000-0003-0407-2266]{Jean-Marc Petit}
\affiliation{Institut UTINAM UMR6213, CNRS, Univ. Bourgogne Franche-Comté, OSU Theta F-25000 Besançon, France}

%% Note that the \and command from previous versions of AASTeX is now
%% depreciated in this version as it is no longer necessary. AASTeX 
%% automatically takes care of all commas and "and"s between authors names.

%% AASTeX 6.3 has the new \collaboration and \nocollaboration commands to
%% provide the collaboration status of a group of authors. These commands 
%% can be used either before or after the list of corresponding authors. The
%% argument for \collaboration is the collaboration identifier. Authors are
%% encouraged to surround collaboration identifiers with ()s. The 
%% \nocollaboration command takes no argument and exists to indicate that
%% the nearby authors are not part of surrounding collaborations.

%% Mark off the abstract in the ``abstract'' environment. 
\begin{abstract}
    
    Dynamically excited objects within the Kuiper belt show a bimodal distribution in their surface colors, and these differing surface colors may be a tracer of where these objects formed. In this work we explore radial color distributions in the primordial planetesimal disk and implications for the positions of ice line/\revision{color} transitions within the Kuiper belt's progenitor populations. We combine a full dynamical model of the Kuiper belt's evolution due to Neptune's migration with precise surface colors measured by the Colours of the Outer Solar System Origins Survey in order to examine the true color ratios within the Kuiper belt and the ice lines within the primordial disk. We investigate the position of a dominant, surface color changing ice-line, with two possible surface color layouts within the initial disk; (1) inner neutral surfaces and outer red, and (2) inner red surfaces and outer neutral. We performed simulations with a primordial disk that truncates at 30 au. By radially stepping the color transition out through 0.5 au intervals we show that both disk configurations are consistent with the observed color fraction. For an inner neutral, outer red primordial disk we find that the color transition can be at $28^{+2}_{-3}$ au at a 95\% confidence level. For an inner red, outer neutral primordial disk the color transition can be at $27^{+3}_{-3}$ au at a 95\% confidence level.
    
\end{abstract}

%% Keywords should appear after the \end{abstract} command. 
%% See the online documentation for the full list of available subject
%% keywords and the rules for their use.

%% From the front matter, we move on to the body of the paper.
%% Sections are demarcated by \section and \subsection, respectively.
%% Observe the use of the LaTeX \label
%% command after the \subsection to give a symbolic KEY to the
%% subsection for cross-referencing in a \ref command.
%% You can use LaTeX's \ref and \label commands to keep track of
%% cross-references to sections, equations, tables, and figures.
%% That way, if you change the order of any elements, LaTeX will
%% automatically renumber them.
%%
%% We recommend that authors also use the natbib \citep
%% and \citet commands to identify citations.  The citations are
%% tied to the reference list via symbolic KEYs. The KEY corresponds
%% to the KEY in the \bibitem in the reference list below. 

\section{Introduction} \label{sec:intro}

    The Kuiper Belt is made up of a sea of icy planetesimals, the remaining relics of planet-forming bodies that failed to evolve into an additional planet in the outer Solar System. Detailed study of these objects sheds light on planetary formation, along with the giant planets' early dynamical history and the compositional structure of the Solar System's primordial planetesimal disk. Kuiper Belt Objects (KBOs) can be split into two broad dynamical classifications: the  dynamically hot population and the dynamically cold classicals. This distinction is due to the dynamical excitement of their orbital properties, the cold classicals reside on low inclination nearly-circular orbits and the hot population has more highly inclined, eccentric orbits. This distinction is consistent with differences in formation location, with the cold classicals having formed at roughly their current positions, and the hot population significantly affected by the migration of Neptune, causing them to be displaced from their formation positions within the Solar System \citep{2005Natur.435..459T,2005Natur.435..462M,2006AJ....131.1142S,2007AJ....133.1962N,2008Icar..196..258L,2010ApJ...722L.204P,2012AJ....144..117N,2015AJ....150...68N}.
    
    The hot population can be split into numerous sub-populations. These include the hot classical KBOs, on close to circular, moderately-to-highly inclined orbits between $\sim39.4$ au and $\sim47.8$ au \citep{2001AJ....121.2804B,gladman_nomenclature_2008}. Resonant KBOs are in mean motion resonances (MMRs) with Neptune. The scattering disk objects are currently scattering off Neptune \citep{1997Sci...276.1670D,1999Icar..141..367P,gladman_nomenclature_2008} and have semimajor axes that extend from $\sim$30 au to $\sim250$ au with perihelion distances 7.35 au $\lesssim q \lesssim$ 40 au. Detached KBOs have \revision{pericenters} decoupled from that of Neptune \citep{2002Icar..157..269G}; perihelion distance $q > 45$ au along with semimajor axis $a > 250$ au \citep{2015MNRAS.446.3788B} is used as the orbit definition in this work. Finally, Centaurs are a short-lived and transitory population, diffusing out of the Kuiper belt \citep{1996Natur.382..507S} with perihelia $q > 7.35$ au, along with Tisserand Parameter ($T_j$) $ > 3.05$ and semimajor axis less than that of Neptune \citep{gladman_nomenclature_2008,morbidelli_kuiper_2019}.
    
    The surface compositions of KBOs can be investigated via either reflectance spectroscopy \citep[e.g.][]{2005A&A...439L...1B, 2007AJ....133..284B, 2007ApJ...670L..49S,2007AJ....133..284B,2015A&A...584A.107B,2015Icar..252..311D} or broadband photometry \citep[e.g.][]{2008ssbn.book...91D,2012ApJ...749...33F,2015A&A...577A..35P,2015ApJ...804...31F,Tegler_Color_2016,pike_col-ossos_2017,2019AJ....157...94M,schwamb_col-ossos:_2019} depending on the size/brightness of the object. Smaller, dimmer KBOs,  with $m_r$ magnitude $>22$, make up the majority of the Kuiper belt population. With fairly featureless spectra, devoid of volatile imprinted features other than water ice \citep{2008ssbn.book..143B}, their surfaces can be studied with the aid of broadband photometry. Surface color variation within the modern day Kuiper belt is important as it provides a window into the Solar System's primordial disk colors \citep{2011ApJ...739L..60B}, enabling the exploration of volatile ice line transitions that triggered color transitions in the early planetary disk. Dynamically excited KBOs are observed to show a bimodal distribution in their surface colors \citep[e.g.][]{1998Natur.392...49T,2003A&A...410L..29P,2012ApJ...749...33F, 2012A&A...546A..86P, 2015A&A...577A..35P, 2015ApJ...804...31F,2017AJ....153..145W,2019AJ....157...94M,schwamb_col-ossos:_2019}.
    
    The objective of this paper is to explore the possible locations of any volatile ice line transitions that may have triggered color variations in the early planetary disk through the combination of dynamical Neptune migration models with Colours of the Outer Solar System Origins Survey \citep[Col-OSSOS,][]{schwamb_col-ossos:_2019} photometry of objects within the modern day Kuiper belt. In Section \ref{sec:colourtranstions} we discuss these observed surface colors in the Kuiper belt, and the proposed primordial disk conditions that caused them. Section \ref{sec:Colour} contains a description of the Col-OSSOS photometry observations. In Section \ref{sec:sampleselection} we describe how we created a comparison sample of Col-OSSOS observations that could later be compared with the color simulations. Section \ref{sec:dynamicalmodel} describes the dynamical model by \citet{nesvorny_neptunes_2016} used in this work. In Section \ref{sec:sims} we explain the color simulations, while Section \ref{sec:results} details the results of these simulations.
    
\section{\revision{Color} Transitions in the Kuiper Belt} \label{sec:colourtranstions}

\begin{figure}
        \centering
        \includegraphics[width=\textwidth]{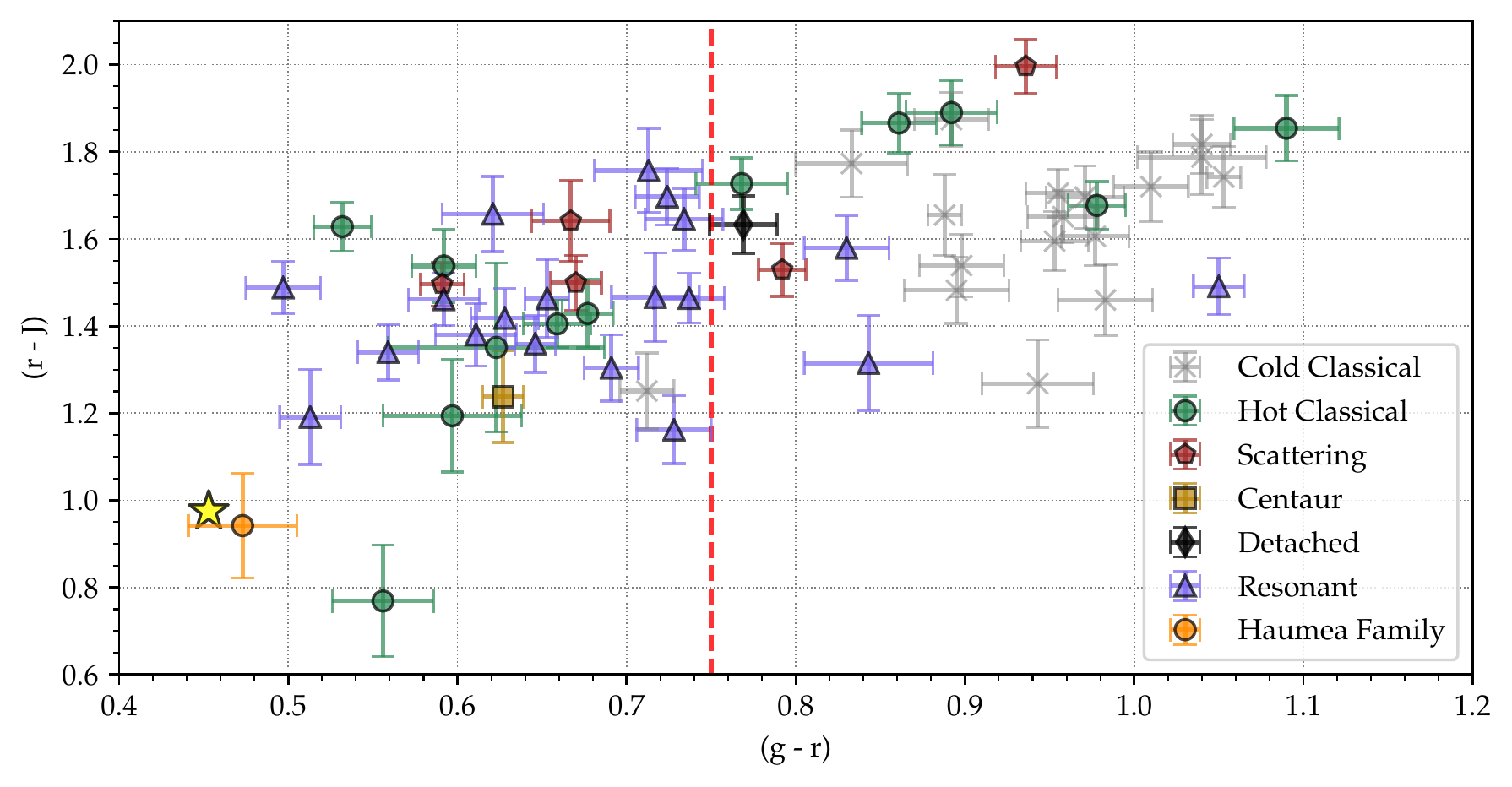}
        \caption{Optical and near-infrared (NIR) colors of Col-OSSOS E, H and L observing blocks. The object 2013 UQ15 (orange circle) is dynamically consistent with the Haumea collisional family. The green circles represent the hot classical KBOs, the \revision{pale \revision{gray}} crosses are the cold classicals, the purple triangles are the MMR objects, the red pentagrams are the scattering disk KBOs, the \revision{black} diamond is detached TNOs and the yellow square is a Centaur. The solar colors, with $g-r=0.45$ and $r-J=0.97$, is shown by the yellow star. \revision{The vertical red dashed line shows the split between red and neutral surface colors at $(g - r) = 0.75$}.}
        \label{fig:colossos_colours}
    \end{figure}

    Observations of the optical/near-infrared colors of non-dwarf planet KBOs reveals a bimodal color distribution \citep[e.g.][]{1998Natur.392...49T,2003A&A...410L..29P,2012ApJ...749...33F, 2012A&A...546A..86P, 2015A&A...577A..35P, 2015ApJ...804...31F,schwamb_col-ossos:_2019}. \revision{In} Figure \ref{fig:colossos_colours} \revision{we plot} the $g-$, $r-$ and $J-$band colors of a sample of KBOs targeted by Col-OSSOS \citep{schwamb_col-ossos:_2019}. Within the color distribution  we \revision{categorize} those KBOs with $(g-r)$ magnitude $\leq0.75$ as `neutral' surfaces, and `red' surfaces as those with $(g -r)$ magnitude $>0.75$. Previous works, such as \citet{2015A&A...577A..35P}, \citet{2017AJ....153..236P} and \citet{schwamb_col-ossos:_2019} have also used this same definition. Cold classical KBOs show optically very red surface colors \citep[e.g.][]{2002ApJ...566L.125T,2005EM&P...97..107L,2008ssbn.book...91D} and thus do not follow the bimodal distribution of the hot population's colors. They are assumed to have formed in place (beyond $\sim40$ au), in different conditions (leading to different surfaces) compared to the dynamically excited Kuiper belt \citep{2010ApJ...722L.204P} and so are not included in this work. The colors shown in Figure \ref{fig:colossos_colours} may hint at possible substructure beyond a simple red/neutral surface classification, with finer transitions such as proposed by \citet{2013Icar..222..307D}. However, we do not have enough resolution on these colors to definitively define any possible substructure. Therefore, as we can distinctly define the red and neutral surfaces we are only exploring this major transition.
    
    There are models in the literature that attempt to explain the differing surface colors for the Kuiper \revision{Belt's} hot population.  These include \citet{2013Icar..222..307D} which presents a taxonomy for KBOs based upon their albedo and colors. \citet{2013Icar..222..307D} suggest that the KBOs can be split into five taxonomic classes and that these taxa show a lack of correlation between their current perihelion distances and their taxonomic properties. Therefore, they suggest that the surface properties are a result of multiple distinct ice-line transitions in the original primordial disk where these objects formed. An alternative model is by \citet{2012ApJ...749...33F}. They suggest that there are three possible surface types in the primordial Kuiper belt; the neutral and red colored surfaces present in the dynamically excited population making up two well defined groups of objects, and the very red colored dynamically cold population making the third. Within this model, the primordial disk originally had two surface types; the cold population surfaces and the dynamically excited KBO surfaces. They suggest that the bifurcation in colors within the hot population is a result of the evolution within the primordial disk. All the objects within this original disk would have started with very similar surface compositions, but based on their positions in the initial disk some of the objects may have lost certain (yet to be determined) volatile species. This resulted in different surface chemistries, and hence, different final surface colors.
    
\begin{figure}
        \centering
        \includegraphics[width=\textwidth]{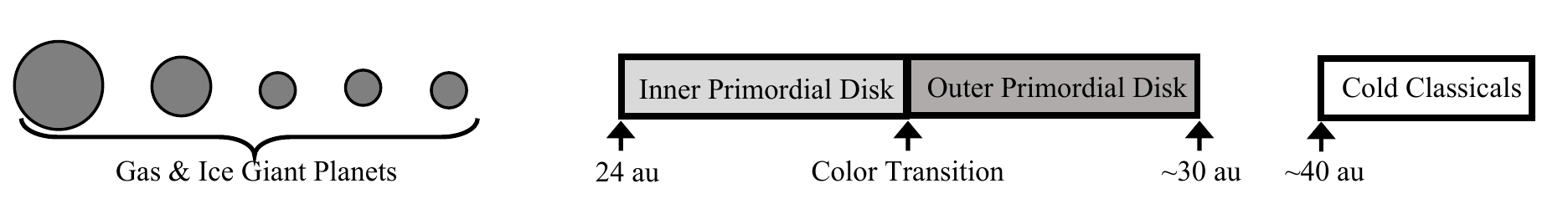}
        \caption{Schematic of the primordial Outer Solar System before giant planet migration. It consists of 2 gas giant and 3 ice giant planets as in the dynamical model of \citet{nesvorny_neptunes_2016}, with Neptune being the outermost planet. The primordial planetesimal disk from which the dynamically hot population originated is represented by the \revision{gray} boxes, and ranges from 24 au to 30 au. It is split into the inner and the outer primordial disk.}
        \label{fig:disk}
    \end{figure}

    In this paper we consider a single dominant surface transition that created the distinct red and neutral surface colors of the dynamically excited objects that we see today. A schematic of this primordial outer Solar System is shown in Figure \ref{fig:disk} and we examine not only the position of the color changing surface transition, but also the order of surface colors in the primordial disk (i.e. an inner neutral / outer red disk vs. an inner red / outer neutral disk). We assumed that this dominant change in surface composition triggered color variations in the early planetary disk, and so divided the red and neutral surfaces found within the dynamically hot population. We are using the combination of an N-body dynamical model of the Kuiper belt through Neptune's migration \citep{nesvorny_neptunes_2016}, along with Col-OSSOS photometry \citep{schwamb_col-ossos:_2019} of objects within the dynamically excited KBOs implanted by Neptune's migration into the Kuiper belt. 
    
    The Col-OSSOS targets were selected to be a brightness complete subsample of the Outer Solar System Origins Survey (OSSOS), which has a well-measured detection efficiency and pointing strategy. This afforded the unique opportunity to explore the true frequency of surface colors within the Kuiper belt. The precision of the colors measured by Col-OSSOS, combined with the well-\revision{characterized} discovery survey \citep[the Outer Solar System Origin Survey (OSSOS),][]{bannister_ossos._2018} enables the accurate investigation of the primordial color distributions. Figures 6 and 7 in \citet{2020AJ....160...46N} show a detailed comparison between OSSOS observations, and a dynamical model biased by the OSSOS survey simulator. The dynamical model of the planetesimal disk throughout Neptune's migration used in this work \citep{nesvorny_neptunes_2016} matches the orbital structure of the Kuiper Belt, while maintaining a precise history of each dynamical test particle. OSSOS made use of the well documented biases and pointing histories inherent in the survey to produce a survey simulator \citep{lawler_ossos:_2018}. This survey simulator allows one to make accurate comparisons between the Col-OSSOS observations and the simulated Kuiper belt from Nesvorný's model, as it can take synthetic planetesimals on simulated orbits and bias them to what OSSOS would have detected.
    
    We test the inner red / outer neutral primordial disk due to the presence of the `blue binaries' within the Kuiper belt \citep{fraser_all_2017}, neutral surfaced wide binaries which simulations suggest are not able to survive a long migration. These are thought to have formed at $\sim$38 au \citep{fraser_all_2017}, separate from both the cold classical KBOs and the dynamically hot KBOs. Here we assume that the neutral surfaced members of the hot population have a similar surface composition to that of the `blue binaries'.  This disk layout was also investigated by \citet{schwamb_col-ossos:_2019}, where they used an analytical model to examine the surface color ratio in the disk. In contrast, we decided to test the inner neutral / outer red disk due to the colors of Neptune's Trojans. Neptune's Trojans, believed to have been captured onto Neptune's orbit during Neptune's migration, have predominantly neutral colored surfaces \citep{2013AJ....145...96P,2018AJ....155...56J}. Additionally the higher inclinations and eccentricities of neutral colored objects \citep{2019AJ....157...94M,2021AJ....162...19A} suggests that they formed closer to the planetary region. This inner neutral / outer red disk is similar to the work of \citet{2020AJ....160...46N} and \citet{2021AJ....162...19A}, where alternative dynamical models of the Kuiper belt throughout Neptune's migration were used to investigate KBO surface colors. In this inner neutral / outer red, if the color transition were far enough out in the disk it could potentially still support the origin scenario for the `blue binaries'.

\section{\revision{Color} Observations}\label{sec:Colour}
    
    The observations used for the primordial color investigations in this paper were taken from Col-OSSOS \citep{schwamb_col-ossos:_2019}. Col-OSSOS selected objects from OSSOS with $r-band$ magnitudes $<$ 23.6 and measured near simultaneous colors of these KBOs in \textit{g-}, \textit{r-} and \textit{J-band} optical and near-infrared filters. They achieved color measurement uncertainties of $\pm$ 0.03 mag for $(g-r)$ for nearly all objects in the sample. OSSOS acquired observations that were grouped into eight regions on the sky called `blocks', each with their own recorded biases and characterisation limits. The Col-OSSOS targets were chosen from six of these eight blocks. For this work we use the H (centered at R.A. 1\textsuperscript{h}35\textsuperscript{m}, decl. +13°28') and L (centered at R.A. 0\textsuperscript{h}54\textsuperscript{m}, decl. +3°50') observing blocks that are published in \citet{schwamb_col-ossos:_2019}, along with E (centered at R.A. 14\textsuperscript{h}50\textsuperscript{m}, decl. -12°32') block \citep[published in][]{wespaper}. E, H and L blocks were used as they provide the most `complete' color sample (minimising the fraction of the sample with unknown colors). Since the release of \citet{schwamb_col-ossos:_2019} some of the Col-OSSOS photometry values have been slightly revised due to minor updates to the software pipeline; these updated values are presented in Table \ref{tab:FullSample}. A complete description of the photometry and data reduction for H and L observing blocks is provided in \citet{schwamb_col-ossos:_2019} and in \citet{wespaper} for E block.
    
\begin{figure}
        \centering
        \includegraphics[width=\textwidth]{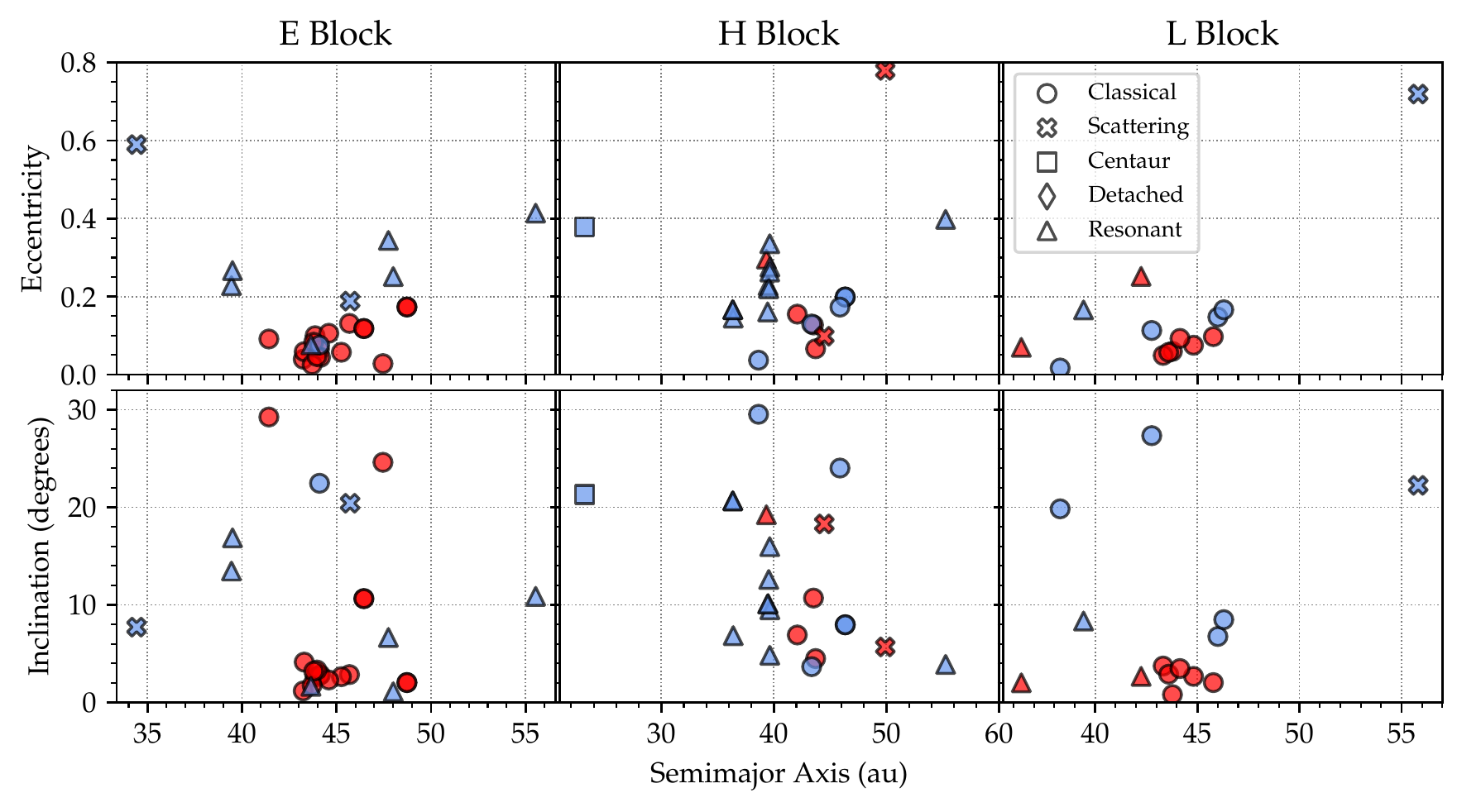}
        \caption{Barycentric orbital parameters, derived from \citet{bannister_ossos._2018}, of TNOs with Col-OSSOS surface colors presented in this paper. One TNO at a$=$150 au in H Block is omitted for better resolution. The 1$\sigma$ uncertainties are smaller than the size of the plot symbol. The colors of the points represent the object's surface colors (as defined in Section \ref{sec:colourtranstions}), with red colors indication red surfaces and blue indicating neutral surfaces.}
        \label{fig:ehlorbitals}
    \end{figure}

    Figure \ref{fig:colossos_colours} shows the $(g-r)$ and $(r-J)$ colors that were measured by Col-OSSOS for these three observing blocks. The resulting bimodal color distribution in the hot population and consistently optically very red surfaces for the cold classicals was in agreement with previous color surveys \citep[e.g.][]{2012ApJ...749...33F, 2012A&A...546A..86P, 2015A&A...577A..35P, 2015ApJ...804...31F,2017AJ....153..236P}. In Table \ref{tab:FullSample} we \revision{summarize} the orbital parameters, along with the colors of the KBOs in the E, H and L block Col-OSSOS sample. In Figure \ref{fig:ehlorbitals} we show the orbital parameters of these KBOs.

    \startlongtable
    \begin{deluxetable}{cccccccccccc}
        \tablecaption{\label{tab:FullSample} Orbital Parameters and Optical and NIR colors of the entire E, H and L Block Col-OSSOS Sample. The KBO classifications are abbreviated with cen = centaur, sca = scattering disk, det = detached, cla = classical belt, $N$:$M$= MMR with Neptune.}
        \tablehead{
            \colhead{MPC} & \colhead{OSSOS ID} & \colhead{Classification} & \colhead{Mean $m_r$} & \colhead{$H_r$} & \colhead{a (au)} & \colhead{e} & \colhead{i ($^{\circ}$)} & \colhead{$(g-r)$} & \colhead{$(r-J)$} }
        \startdata
        2002 GG166 &    o3e01 &  sca &    21.5$\pm$0.09 &   7.73 &   34.42 &  0.590 &   7.71 &   0.59$\pm$0.01 &   1.5$\pm$0.05 \\
            2013 GH137 &    o3e02 &  3:2 &   23.34$\pm$0.14 &   8.32 &   39.44 &  0.228 &  13.47 &   0.71$\pm$0.03 &   1.76$\pm$0.1 \\
            2013 GJ137 &    o3e04 &  3:2 &   23.39$\pm$0.16 &   8.25 &   39.50 &  0.267 &  16.87 &   0.62$\pm$0.03 &  1.66$\pm$0.09 \\
            2013 GW136 &    o3e05 &  2:1 &   22.69$\pm$0.07 &   7.42 &   47.74 &  0.344 &   6.66 &   0.72$\pm$0.02 &   1.7$\pm$0.06 \\
            2013 GY136 &    o3e09 &  5:2 &   22.94$\pm$0.05 &   7.32 &   55.54 &  0.414 &  10.88 &   0.51$\pm$0.02 &  1.48$\pm$0.07 \\
            2013 GS137 &    o3e16 &  cla &   23.47$\pm$0.14 &   7.44 &   43.87 &  0.100 &   2.60 &   1.01$\pm$0.02 &  1.72$\pm$0.08 \\
            2013 GR136 &    o3e19 &  7:4 &     23.4$\pm$0.1 &   7.20 &   43.65 &  0.076 &   1.64 &   0.72$\pm$0.03 &   1.47$\pm$0.1 \\
            2001 FK185 &  o3e20PD &  cla &   23.09$\pm$0.22 &   6.82 &   43.24 &  0.039 &   1.17 &   0.83$\pm$0.03 &  1.77$\pm$0.08 \\
            2013 GQ137 &    o3e21 &  cla &    23.4$\pm$0.09 &   7.12 &   45.69 &  0.131 &   2.85 &   0.89$\pm$0.02 &  1.87$\pm$0.06 \\
            2013 GN137 &    o3e22 &  cla &   22.97$\pm$0.09 &   6.70 &   44.09 &  0.065 &   2.76 &   1.05$\pm$0.01 &  1.74$\pm$0.07 \\
            2001 FO185 &  o3e23PD &  cla &   23.37$\pm$0.08 &   7.09 &   46.45 &  0.118 &  10.64 &   0.86$\pm$0.02 &  1.87$\pm$0.07 \\
             2004 EU95$^{1}$ &  o3e27PD &  cla &     23.1$\pm$0.1 &   6.77 &   44.15 &  0.044 &   2.82 &   0.97$\pm$0.02 &    ...$\pm$... \\
            2013 GX137 &    o3e28 &  cla &    23.17$\pm$0.1 &   6.82 &   43.29 &  0.059 &   4.13 &   0.98$\pm$0.03 &  1.46$\pm$0.08 \\
            2013 GO137 &    o3e29 &  cla &   23.46$\pm$0.08 &   7.09 &   41.42 &  0.091 &  29.25 &   0.77$\pm$0.03 &  1.73$\pm$0.06 \\
            2013 EM149 &  o3e30PD &  cla &   22.99$\pm$0.05 &   6.59 &   45.26 &  0.057 &   2.63 &   0.96$\pm$0.02 &  1.65$\pm$0.06 \\
            2013 GT137 &    o3e31 &  cla &   23.55$\pm$0.13 &   7.10 &   44.59 &  0.106 &   2.29 &   1.04$\pm$0.04 &  1.79$\pm$0.09 \\
            2013 GF138 &  o3e34PD &  cla &    23.57$\pm$0.1 &   7.05 &   44.04 &  0.019 &   0.55 &   1.07$\pm$0.03 &  1.71$\pm$0.06 \\
            2013 GP137 &    o3e35 &  cla &   23.48$\pm$0.13 &   6.94 &   43.71 &  0.025 &   1.75 &   0.94$\pm$0.03 &   1.27$\pm$0.1 \\
             2004 HJ79 &  o3e37PD &  cla &   23.37$\pm$0.09 &   6.81 &   43.96 &  0.046 &   3.32 &   0.95$\pm$0.02 &   1.6$\pm$0.07 \\
            2013 GP136 &    o3e39 &  det &   23.07$\pm$0.07 &   6.42 &  150.24 &  0.727 &  33.54 &   0.77$\pm$0.02 &  1.63$\pm$0.07 \\
            2013 GV137$^{1}$ &    o3e43 &  cla &   23.42$\pm$0.28 &   6.67 &   43.79 &  0.083 &   3.20 &   0.95$\pm$0.06 &    ...$\pm$... \\
            2013 GG138 &    o3e44 &  cla &   23.26$\pm$0.09 &   6.34 &   47.46 &  0.028 &  24.61 &   1.09$\pm$0.03 &  1.85$\pm$0.08 \\
            2013 GQ136$^{1}$ &    o3e45 &  cla &    23.59$\pm$0.1 &   6.13 &   48.73 &  0.173 &   2.03 &   1.08$\pm$0.02 &    ...$\pm$... \\
            2013 HR156$^{1}$ &    o3e49 & 15:8 &   23.54$\pm$0.09 &   7.72 &   45.72 &  0.188 &  20.41 &   0.58$\pm$0.03 &    ...$\pm$... \\
            2013 GM137 &    o3e51 &  cla &   23.32$\pm$0.23 &   6.90 &   44.10 &  0.076 &  22.46 &    0.6$\pm$0.04 &  1.19$\pm$0.13 \\
            2013 GX136 &    o3e55 &  2:1 &   23.41$\pm$0.13 &   7.67 &   48.00 &  0.252 &   1.10 &   0.73$\pm$0.02 &  1.64$\pm$0.07 \\
             2013 UR15 &    o3l01 &  sca &   23.06$\pm$0.06 &  10.89 &   55.82 &  0.719 &  22.25 &   0.67$\pm$0.02 &  1.64$\pm$0.09 \\
            2001 QF331 &  o3l06PD &  5:3 &   22.71$\pm$0.07 &   7.56 &   42.25 &  0.252 &   2.67 &   0.83$\pm$0.02 &  1.58$\pm$0.07 \\
             2013 US15 &    o3l09 &  4:3 &   23.24$\pm$0.16 &   7.78 &   36.38 &  0.070 &   2.02 &   1.05$\pm$0.02 &  1.49$\pm$0.06 \\
            2003 SR317 &  o3l13PD &  3:2 &   23.36$\pm$0.08 &   7.66 &   39.43 &  0.166 &   8.35 &   0.65$\pm$0.01 &  1.36$\pm$0.06 \\
             2013 SZ99 &    o3l15 &  cla &   23.54$\pm$0.13 &   7.65 &   38.28 &  0.017 &  19.84 &   0.59$\pm$0.02 &  1.54$\pm$0.08 \\
            2010 RE188 &    o3l18 &  cla &   22.27$\pm$0.05 &   6.19 &   46.01 &  0.147 &   6.76 &   0.68$\pm$0.02 &  1.43$\pm$0.08 \\
             2013 SP99 &    o3l32 &  cla &   23.47$\pm$0.08 &   7.23 &   43.78 &  0.060 &   0.79 &   0.98$\pm$0.02 &  1.61$\pm$0.07 \\
             2013 UL15 &    o3l43 &  cla &   23.05$\pm$0.11 &   6.62 &   45.79 &  0.097 &   2.02 &    0.9$\pm$0.03 &  1.48$\pm$0.08 \\
             2013 UO15 &    o3l50 &  cla &    23.2$\pm$0.06 &   6.69 &   43.33 &  0.049 &   3.73 &   0.96$\pm$0.02 &   1.7$\pm$0.06 \\
            2006 QF181 &    o3l60 &  cla &   23.29$\pm$0.07 &   6.79 &   44.81 &  0.075 &   2.66 &    0.9$\pm$0.02 &  1.54$\pm$0.07 \\
             2013 UX18 &    o3l69 &  cla &    23.42$\pm$0.1 &   6.74 &   43.60 &  0.057 &   2.89 &   0.89$\pm$0.01 &  1.66$\pm$0.09 \\
             2013 SQ99 &    o3l76 &  cla &    23.1$\pm$0.06 &   6.35 &   44.15 &  0.093 &   3.47 &   0.97$\pm$0.02 &   1.7$\pm$0.07 \\
             2013 UQ15 &    o3l77 &  cla &   22.93$\pm$0.12 &   6.07 &   42.77 &  0.113 &  27.34 &   0.47$\pm$0.03 &  0.94$\pm$0.12 \\
            2013 SA100 &    o3l79 &  cla &   22.81$\pm$0.04 &   5.77 &   46.30 &  0.166 &   8.48 &   0.66$\pm$0.02 &   1.4$\pm$0.05 \\
            2014 UJ225 &    o4h01 &  cen &   22.74$\pm$0.12 &  10.29 &   23.20 &  0.378 &  21.32 &   0.63$\pm$0.01 &   1.24$\pm$0.1 \\
            2014 UQ229 &    o4h03 &  sca &   22.69$\pm$0.21 &   9.55 &   49.90 &  0.779 &   5.68 &   0.94$\pm$0.02 &   2.0$\pm$0.06 \\
            2014 UX229 &    o4h05 &  3:2 &   22.25$\pm$0.04 &   8.04 &   39.63 &  0.335 &  15.97 &   0.65$\pm$0.01 &  1.46$\pm$0.09 \\
            2010 TJ182 &    o4h07 &  3:2 &   22.28$\pm$0.02 &   7.68 &   39.65 &  0.276 &   9.50 &   0.56$\pm$0.02 &  1.34$\pm$0.06 \\
            2014 UV228 &    o4h09 &  3:2 &   23.48$\pm$0.08 &   8.49 &   39.49 &  0.228 &  10.14 &   0.59$\pm$0.02 &  1.46$\pm$0.06 \\
            2014 UO229 &    o4h11 &  3:2 &   23.55$\pm$0.07 &   8.25 &   39.45 &  0.161 &  10.09 &   0.73$\pm$0.02 &  1.16$\pm$0.08 \\
            2014 UD229 &    o4h13 &  4:3 &   23.54$\pm$0.07 &   8.18 &   36.39 &  0.145 &   6.85 &   0.69$\pm$0.02 &   1.3$\pm$0.08 \\
            2014 US229 &    o4h14 &  5:2 &   23.18$\pm$0.08 &   7.95 &   55.26 &  0.398 &   3.90 &   0.63$\pm$0.02 &  1.42$\pm$0.07 \\
            2014 UX228 &    o4h18 &  4:3 &   23.11$\pm$0.05 &   7.35 &   36.35 &  0.167 &  20.66 &    0.5$\pm$0.02 &  1.49$\pm$0.06 \\
            2014 UK225 &    o4h19 &  cla &   23.23$\pm$0.06 &   7.43 &   43.52 &  0.127 &  10.69 &   0.98$\pm$0.02 &  1.68$\pm$0.06 \\
            2014 UL225 &    o4h20 &  cla &   23.03$\pm$0.07 &   7.24 &   46.34 &  0.199 &   7.95 &   0.56$\pm$0.03 &  0.77$\pm$0.13 \\
            2014 UH225 &    o4h29 &  cla &   23.31$\pm$0.06 &   7.30 &   38.64 &  0.037 &  29.53 &   0.53$\pm$0.02 &  1.63$\pm$0.06 \\
            2014 UM225 &    o4h31 &  9:5 &   23.25$\pm$0.06 &   7.21 &   44.48 &  0.098 &  18.30 &   0.79$\pm$0.01 &  1.53$\pm$0.06 \\
            2007 TC434 &    o4h39 &  9:1 &   23.21$\pm$0.05 &   7.13 &  129.94 &  0.695 &  26.47 &   0.67$\pm$0.02 &   1.5$\pm$0.06 \\
            2014 UD225 &    o4h45 &  cla &   23.09$\pm$0.05 &   6.63 &   43.36 &  0.130 &   3.66 &   0.71$\pm$0.02 &  1.25$\pm$0.09 \\
            2001 RY143 &    o4h48 &  cla &   23.54$\pm$0.08 &   6.80 &   42.08 &  0.155 &   6.91 &   0.89$\pm$0.03 &  1.89$\pm$0.07 \\
            2014 UE225 &    o4h50 &  cla &   22.67$\pm$0.04 &   5.99 &   43.71 &  0.066 &   4.49 &   1.04$\pm$0.02 &  1.82$\pm$0.07 \\
              1995 QY9 &  o4h69PD &  3:2 &   22.38$\pm$0.06 &   7.68 &   39.64 &  0.263 &   4.84 &   0.74$\pm$0.02 &  1.46$\pm$0.06 \\
            2014 UF228 &    o4h70 &  3:2 &    22.7$\pm$0.04 &   7.77 &   39.55 &  0.220 &  12.60 &   0.61$\pm$0.02 &  1.38$\pm$0.07 \\
            2001 RX143 &  o4h76PD &  3:2 &   22.84$\pm$0.06 &   6.42 &   39.34 &  0.296 &  19.24 &   0.84$\pm$0.04 &  1.32$\pm$0.11 \\
          \enddata
          \tablenotetext{^1}{These objects do not have J-band observations.}
          \tablenotetext{}{\textbf{N.B.} The third character in the OSSOS ID denotes the discovery block of the object.}
    \end{deluxetable}
    
    \subsection{Potential Correlation in Neutral Class}
        
        \citet{schwamb_col-ossos:_2019} report a tentative anti-correlation in the neutral colored objects (those KBOs with $(g-r) < 0.75$) of their sample in $(g-r)/(r-J)$ color space. We performed a Spearman rank statistic test on the colors of our E, H and L block targets. We removed the Haumea collisional family member (2013 UQ15) from the color sample for this statistical test due to the object's surface having been produced via collision \citep{2008ApJ...684L.107S,2010A&A...511A..72S,2011ApJ...730..105T,2012A&A...544A.137C,2012ApJ...749...33F,2020NatAs...4...89P}. The Spearman rank test found no significant correlation, with a correlation coefficient of 0.027 and an 89.4\% probability that any correlation occurred by chance. We also performed this test with the potential outlier 2014 UL225 removed from the sample, similarly to \citet{schwamb_col-ossos:_2019}. This object has a significantly different surface color to the rest of the neutral cloud (with $(g-r) = 0.56\pm0.03$ and $(r-J) =  0.77\pm0.13$). In this case we found no significant correlation again, with a correlation coefficient of -0.085 and a 68.7\% chance that this occurred by chance. Therefore, we find that there is no evidence for an anti-correlation in $(g-r)/(r-J)$ color space.
        
\section{Col-OSSOS Comparison Sample} \label{sec:sampleselection}
    
    The Col-OSSOS observations outlined in Section \ref{sec:Colour} could not be directly compared with the simulation outputs we discuss in Section \ref{sec:modeldynamicalcuts}. For any comparisons between the two to be accurate we had to ensure consistent treatment of both the observed and the simulated KBOs. Here we \revision{summarize} the creation of a subsample of Col-OSSOS observations that could be compared with the simulations, and refer hereafter to this subsample as the `comparison sample'.

    \subsection{Observational Limits}\label{sec:coloursample}
        
        The OSSOS survey simulator \citep{lawler_ossos:_2018} was used to bias the dynamical model using the pointings and detection limits of OSSOS. This allowed accurate comparison between the simulated Kuiper belt and the comparison sample. The OSSOS survey simulator uses the mean discovery magnitudes of the simulated KBOs to decide what would have been detected. Col-OSSOS selected any OSSOS targets with magnitudes brighter than $H_r$ of 23.6 to observe for color studies, and this target selection was occurring while OSSOS was still finding new targets. However, due to a re-calibration of the OSSOS photometry in 2014, the initial magnitudes of some of the targets shifted. Due to this, the OSSOS target 2013 UM17 (in L observing block) with a discovery magnitude of 23.56 was not selected as a Col-OSSOS target, as its preliminary magnitude did not make the Col-OSSOS cutoff. Therefore, we include this object in our comparison sample and classify its surface color as `unknown'.

        As mentioned in Section \ref{sec:colourtranstions}, we split the Col-OSSOS colors into red and neutral surfaces based on their $(g - r)$ magnitude (red surfaces with $(g -r)>0.75$ and neutral surfaces with $(g -r)\leq0.75$). There was one object in E block (2013 HR156) that did not have $J-band$ observations due to an incomplete observation. However, as our red/neutral split was based entirely on the $(g - r)$ magnitude the color of this object could still be \revision{characterized} and so it was kept within the comparison sample.
        
    \subsection{Haumea Collisional Family}
        
        2013 UQ15 was a Col-OSSOS observed member of the Haumea collisional family \citep{2020NatAs...4...89P}. These \revision{are} collisional fragments of the dwarf planet Haumea created in a long ago collision, and are distinguished by strong water ice absorption on their surfaces and clustered orbital properties, along with neutral optical surface colors \citep{2007AJ....133..284B,2008ApJ...684L.107S,2010A&A...511A..72S,2011ApJ...730..105T,2012A&A...544A.137C,2012ApJ...749...33F,2019AJ....157..230P}. Therefore, as the surfaces and orbits within this family are not primordial, 2013 UQ15 is removed from the comparison sample.
        
    \subsection{Dynamical Cuts} \label{sec:dynamical}
    
        In this work we were investigating the hot population KBOs that were implanted onto their current orbits by interactions with Neptune. In order to investigate the red and neutral surfaces of the hot population KBOs, we wanted \revision{our} sample made up of those KBOs that have hot population surfaces and were emplaced onto their orbits by Neptune's migration. \revision{Therefore, we made a series of dynamical cuts to our Col-OSSOS comparison sample. The resulting sample was made up of Col-OSSOS KBOs on hot classical, scattering and detached orbits.}
        
        We ensured that there were no KBOs on cold classical orbits in either the comparison sample or in the simulated KBO population. Although cold classical KBOs are generally defined as those objects with orbital inclination less than 5 degrees, the tail of the hot population inclination distribution, and other dynamical classes such as the MMR KBOs and scattering disk objects, overlap. OSSOS strove to dynamically classify their observed KBOs accurately. However, as the dynamical model did not classify the synthetic KBOs dynamical classes, we chose to \revision{use} the definition of classical orbits from \citet{gladman_nomenclature_2008}, giving an approximate definition with semimajor axes between $37.37$ au $< a < 55.1$ au, and eccentricities less than 0.24. This allowed us to define cold classical KBOs as those on classical orbits with inclinations less than 5$^{\circ}$, and therefore remove all OSSOS defined low inclination classical KBOs. By applying this definition we were also able to remove any simulated KBOs on the same orbits, as described in Section \ref{sec:modeldynamicalcuts}.
        
        Objects in strong MMRs are likely to include objects with cold classical surfaces \revision{captured} during planetary migration \citep{2019AJ....158...53T}, so removing objects trapped in MMRs is necessary to understand the color distribution of the dynamically excited sample. As part of the OSSOS sample, our targets have multi-year arcs and careful classification as resonant/nonresonan \citep{2016AJ....152...23V}. We chose to remove KBOs in the main MMRs \citep[3:2, 5:2, 4:3, 5:3, 7:4 and 2:1,][]{nesvorny_neptunes_2016} from our comparison sample. Within OSSOS MMRs were identified up to very high orders, such as the 15:9 MMR. As these high order MMRs have inclination distributions consistent with the hot classicals, and did not sweep through the cold classical region during the migration period, we chose to group these KBOs with the scattering population. Therefore, seven objects on these high order resonances have been redefined as scattering and included in our comparison sample.
        
        Due to the short lifetimes of Centaurs, the Centaur within Col-OSSOS (2014 UJ225) likely diffused onto its orbit much later than the end of Neptune's migrations. Therefore, we could not accurately infer where in the primordial disk any Centaurs originated. Along with this, due to Centaur's orbits being closer in to the Sun than other KBOs their surfaces may undergo thermal processing that would change their surface colors. We adapted a definition for Centaurs from \citet{gladman_nomenclature_2008}, and used semimajor axis less than that of Neptune, aphelion distance $>$ 11 au as this definition removed our only single Col-OSSOS centaur from the sample and could also be applied to the dynamical model sample. For similar reasons, any objects with semimajor axis greater than 250 au, and perihelion distance $<45$ au were removed from the sample as their orbits have likely significantly evolved since the end of Neptune's migration \citep{2015MNRAS.446.3788B}.
        
    \subsection{Comparison Sample Statistics}

\begin{figure}
            \centering
            \includegraphics[width=\textwidth]{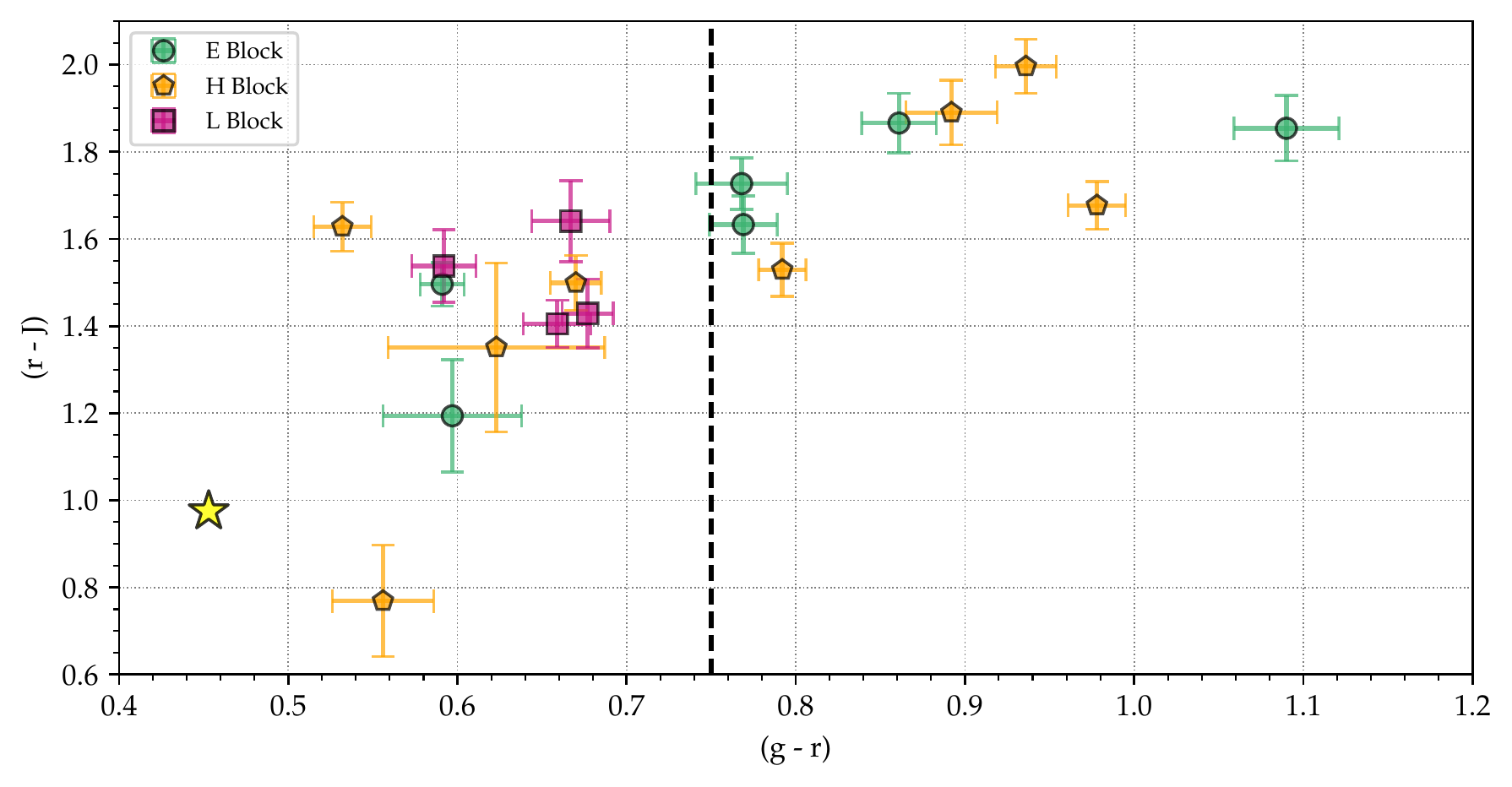}
            \caption{Col-OSSOS photometry of observed non-resonant, non-Centaur, dynamically excited objects in E, H and L observing blocks. The red/neutral color split is placed at a $(g - r)$ magnitude of 0.75. The color distribution of objects before these cuts in E, H and L blocks is shown in Figure \ref{fig:colossos_colours}. The star shows solar colors.}
            \label{fig:sample_colours}
        \end{figure}

\begin{figure}
            \centering
            \includegraphics[width=\textwidth]{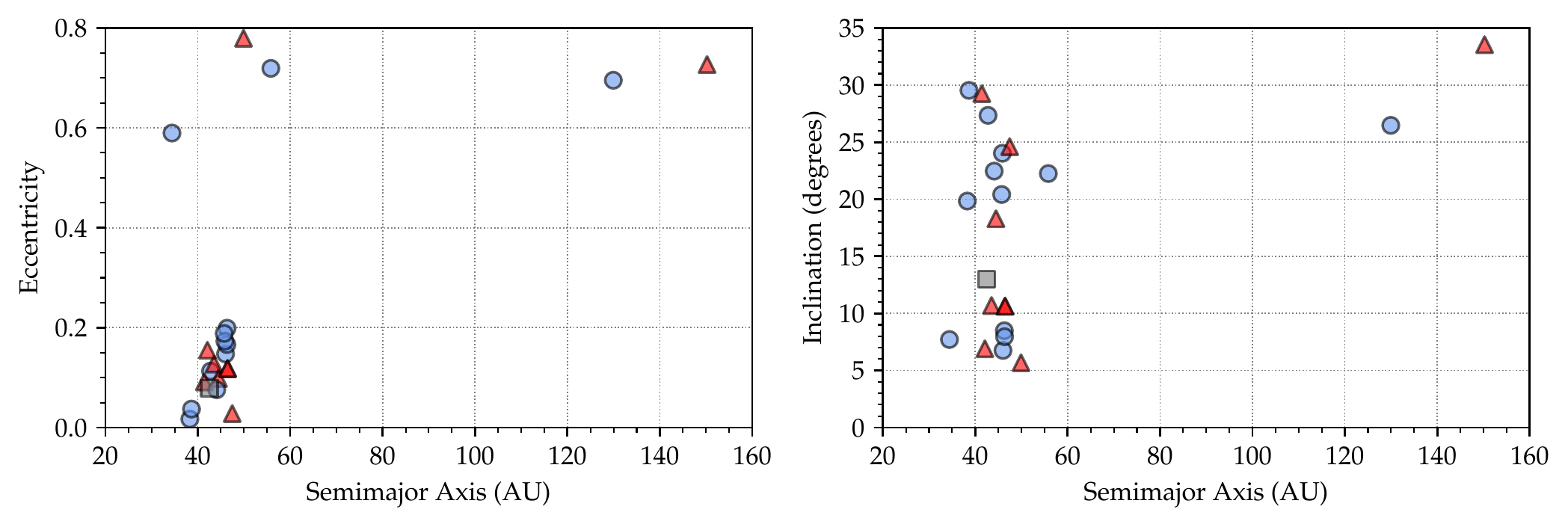}
            \caption{\revision{Barycentric orbital parameters, derived from \citet{bannister_ossos._2018}, of observed non-resonant, non-Centaur, dynamically excited objects in E, H and L observing blocks. Blue circles represent those objects with neutral colored surfaces ($(g - r) < 0.75$), red triangles are the objects with red colored surfaces ($(g - r) \geq 0.75$) and the \revision{gray} square represents the KBO within the sample with no Col-OSSOS observations.}}
            \label{fig:sample_orbits}
        \end{figure}

        After the various limits outlined above we were left with 20 KBOs, making up our comparison sample of Col-OSSOS observations. The resulting sample is shown in Table \ref{tab:sample}, including the red/neutral color class assigned based on their $(g - r)$ magnitude as described in Section \ref{sec:coloursample}. It is made up of 13 hot classical KBOs, 6 scattering KBOs and one detached KBO. In Figure \ref{fig:sample_colours} we show the observed $g-$, $r-$, and $J-$band colors of the sample, along with their corresponding observing block. \revision{In Figure \ref{fig:sample_orbits} we plot the orbital elements of the comparison sample objects.} This left us with a total of 9 red surfaced KBOs, 10 with neutral surfaces and 1 object with unknown surface colors. These numbers are \revision{summarized} in Table \ref{tab:colours_summary}.
        
        \begin{deluxetable}{cccccccccccc}
            \tablecaption{\label{tab:sample} Orbital Parameters and Optical and NIR Colors of the comparison KBO Sample. The KBO classifications are abbreviated with cen = centaur, sca = scattering disk, det = detached, cla = classical belt, $N$:$M$= MMR with Neptune.}
            \tablehead{
                \colhead{MPC} & \colhead{OSSOS ID} & \colhead{Classification} & \colhead{Mean $m_r$} & \colhead{$H_r$} & \colhead{a (au)} & \colhead{e} & \colhead{i ($^{\circ}$)} & \colhead{$(g-r)$} & \colhead{$(r-J)$} }
            \startdata
                 2002 GG166 &    o3e01 &    sca &  21.5$\pm$0.09 &   7.73 &   34.42 &  0.590 &   7.71 &  0.59$\pm$0.01 &   1.5$\pm$0.05 \\
                 2001 FO185 &  o3e23PD &    cla &  23.37$\pm$0.08 &   7.09 &   46.45 &  0.118 &  10.64 & 0.86$\pm$0.02 &  1.87$\pm$0.07 \\
                 2013 GO137 &    o3e29 &    cla &  23.46$\pm$0.08 &   7.09 &   41.42 &  0.091 &  29.25 & 0.77$\pm$0.03 &  1.73$\pm$0.06 \\
                 2013 GP136 &    o3e39 &    det &  23.07$\pm$0.07 &   6.42 &  150.24 &  0.727 &  33.54 & 0.77$\pm$0.02 &  1.63$\pm$0.07 \\
                 2013 GG138 &    o3e44 &    cla &  23.26$\pm$0.09 &   6.34 &   47.46 &  0.028 &  24.61 & 1.09$\pm$0.03 &  1.85$\pm$0.08 \\
                 2013 HR156 &    o3e49 &    sca &  23.54$\pm$0.09 &   7.72 &   45.72 &  0.188 &  20.41 & 0.58$\pm$0.03 &    ...$\pm$... \\
                 2013 GM137 &    o3e51 &   cla  &  23.32$\pm$0.23 &   6.90 &   44.10 &  0.076 &  22.46 &  0.6$\pm$0.04 &  1.19$\pm$0.13 \\
                 2013 UR15  &    o3l01 &    sca &  23.06$\pm$0.06 &  10.89 &   55.82 &  0.719 &  22.25 & 0.67$\pm$0.02 &  1.64$\pm$0.09 \\
                 2013 SZ99  &    o3l15 &    cla &  23.54$\pm$0.13 &   7.65 &   38.28 &  0.017 &  19.84 & 0.59$\pm$0.02 &  1.54$\pm$0.08 \\
                 2010 RE188 &    o3l18 &    cla &  22.27$\pm$0.05 &   6.19 &   46.01 &  0.147 &   6.76 & 0.68$\pm$0.02 &  1.43$\pm$0.08 \\
                 2013 UM17$^{1}$&o3l29PD &  cla &  23.56$\pm$0.09 &   7.29 &   42.48 &  0.079 &  12.99 &  ...$\pm$...  &  ...$\pm$...   \\
                %  2013 UQ15  &    o3l77 &    cla &  22.93$\pm$0.12 &   6.07 &   42.77 &  0.113 &  27.34 & 0.47$\pm$0.03 &  0.94$\pm$0.12 \\
                 2013 SA100 &    o3l79 &    cla &  22.81$\pm$0.04 &   5.77 &   46.30 &  0.166 &   8.48 & 0.66$\pm$0.02 &   1.4$\pm$0.05 \\
                 2014 UQ229 &    o4h03 &    sca &  22.69$\pm$0.21 &   9.55 &   49.90 &  0.779 &   5.68 & 0.94$\pm$0.02 &   2.0$\pm$0.06 \\
                 2014 UK225 &    o4h19 &    cla &  23.23$\pm$0.06 &   7.43 &   43.52 &  0.127 &  10.69 & 0.98$\pm$0.02 &  1.68$\pm$0.06 \\
                 2014 UL225 &    o4h20 &    cla &  23.03$\pm$0.07 &   7.24 &   46.34 &  0.199 &   7.95 & 0.56$\pm$0.03 &  0.77$\pm$0.13 \\
                 2014 UH225 &    o4h29 &    cla &  23.31$\pm$0.06 &   7.30 &   38.64 &  0.037 &  29.53 & 0.53$\pm$0.02 &  1.63$\pm$0.06 \\
                 2014 UM225 &    o4h31 &    sca &  23.25$\pm$0.06 &   7.21 &   44.48 &  0.098 &  18.30 & 0.79$\pm$0.01 &  1.53$\pm$0.06 \\
                 2007 TC434 &    o4h39 &    sca &  23.21$\pm$0.05 &   7.13 &  129.94 &  0.695 &  26.47 & 0.67$\pm$0.02 &   1.5$\pm$0.06 \\
                 2001 RY143 &    o4h48 &    cla &  23.54$\pm$0.08 &   6.80 &   42.08 &  0.155 &   6.91 & 0.89$\pm$0.03 &  1.89$\pm$0.07 \\
                 2014 UN228 &    o4h75&    cla &  23.37$\pm$0.11 &   7.46 &   45.87 &  0.173 &  24.02 &  0.62$\pm$0.06 	  &  1.35$\pm$0.19   \\
            \enddata
            \tablenotetext{^1}{This object does not have Col-OSSOS observations.}
        \end{deluxetable}
            
        \begin{deluxetable}{c|ccc}
            \tablecaption{\label{tab:colours_summary} Summary of the numbers of each surface color in each observing block in the resulting comparison sample.}
            \tablehead{
                \colhead{Observing Block}  & \colhead{Neutral} & \colhead{Red} & \colhead{Unknown} }
                \startdata
                E     & 3   & 4    & 0    \\
                H     & 4   & 4    & 0    \\
                L     & 4   & 0    & 1    \\
                \hline
                Total & 11  & 8    & 1    \\
              \enddata
        \end{deluxetable}
        
\section{Dynamical Model of Neptune's Migration} \label{sec:dynamicalmodel}
    
    In order to create our simulated Kuiper belt population we took the orbital parameters of synthetic KBOs from \citet{nesvorny_neptunes_2016}, and applied a color transition to the pre-Neptune migration disk. The 5 planet migration model used a slow migration of Neptune from 20 au to $\sim$30 au with a `jump', or sudden change in semimajor axis at 28 au along with massive planetesimals in the primordial disk for Neptune to scatter off. \revision{The slow migration introduced in \citet{2015AJ....150...68N} widened the inclination distribution of the resulting Kuiper belt, while Neptune's migration `jump' in \citet{2015AJ....150...73N} aids in the creation of the cold classical kernel.} The addition of massive planetesimals to the disk in \citet{nesvorny_neptunes_2016} caused Neptune's migration to be `grainy', destabilising the resonant bodies that have large libration amplitudes and causing them to end up on stable non-resonant orbits. They found that they could achieve a best fit for the Canada–France Ecliptic Plane Survey \citep[CFEPS,][]{2009AJ....137.4917K, 2011AJ....142..131P} observations they compared with, when their simulations contained 1000 - 4000 Pluto-sized objects in their pre-Neptune migration disk. This model of Neptune's migration well matched the known Kuiper belt, while maintaining a precise history of each dynamical test particle.
    
\begin{figure}
        \centering
        \includegraphics[width=\textwidth]{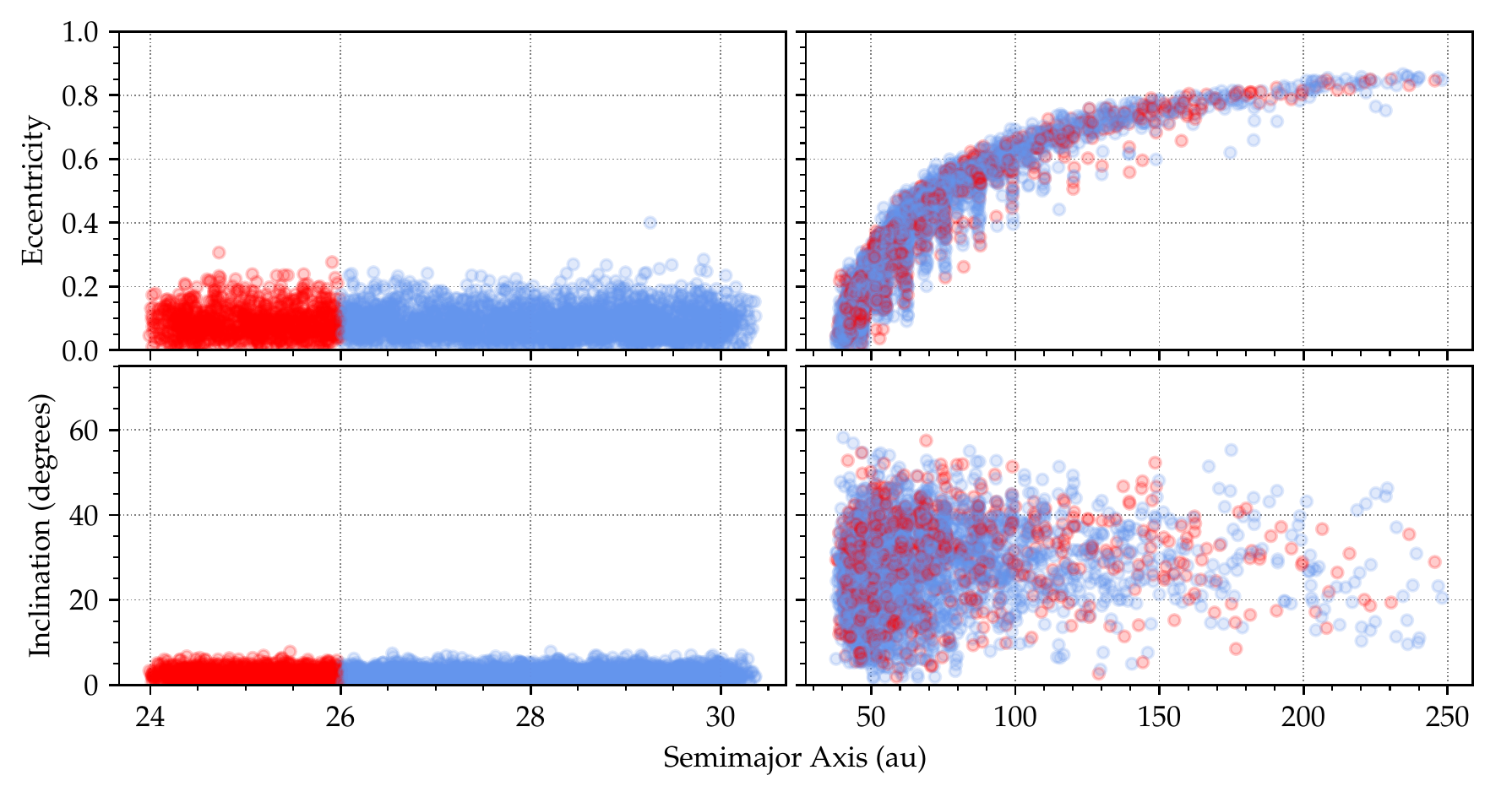}
        \caption{An example simulated Kuiper belt population from the model of \citet{nesvorny_neptunes_2016} after the dynamical cuts outlined in Section \ref{sec:modeldynamicalcuts}. This example has an inner red / outer neutral primordial disk, and color transition at 26 au. The red points represent redder surfaced KBOs, while the blue points represent neutral surfaced KBOs. The left column shows the objects in the primordial disk, while the right column show the objects post-Neptune migration from the model of \citet{nesvorny_neptunes_2016}.}
        \label{fig:nes_colours}
    \end{figure}

    Figure \ref{fig:nes_colours} shows the pre- and post-Neptune migration disk for this model, with the synthetic KBOs colored based on their position in the primordial disk. The post-Neptune migration disk is from immediately after the end of Neptune's migration, \revision{it} shows the synthetic particles that survive Neptune's migration with no further integration. \revision{While we account for precession effects over the 4 Gyrs past Neptune's migration, dynamical erosion over this time period could potentially have a small impact the fractions of red and neutral objects that remain in the present day. However, as this would impact both the red and neutral colored populations similarly it does dot have a significant impact on our final result.} Each individual test particle in the dynamical model was precisely tracked, and so the orbital parameters of the simulated KBOs immediately after the period of giant planet migration were recorded. We could therefore create a synthetic Kuiper belt hot population with colors that corresponded to the original locations of these objects within the primordial Kuiper belt.
    
    \subsection{Dynamical Cuts} \label{sec:modeldynamicalcuts}
        
        As mentioned in Section \ref{sec:dynamical}, dynamical cuts were applied to the observed sample to create a comparison sample of observations. In order to make accurate comparisons between this comparison sample and the simulation data, both had to undergo the same treatment. Therefore we applied the same dynamical cuts outlined in Section \ref{sec:dynamical} to the synthetic post-Neptune migration disk. We removed the major MMRs identified by \citet{nesvorny_neptunes_2016} (3:2, 5:2, 4:3, 5:3, 7:4 and 2:1). Although the dynamical model only consisted of the dynamically hot population, we removed any classical defined synthetic KBOs using the same conditions as Section \ref{sec:dynamical}. Similarly we also removed Centaur type orbits (semimajor axis less than that of Neptune, aphelion distance $>$ 11 au), along with objects with semimajor axis $>250$ au and perihelion distance $<45$ au. After the dynamical cuts, the post-Neptune migration disk consisted of $\sim$3500 synthetic KBOs out of the original $\sim$4200.
        
\section{\revision{Color} Simulations} \label{sec:sims}

    The aim of this work is to investigate the radial distribution of red and neutral surfaced objects in the pre-Neptune migration disk. As outlined in Section \ref{sec:colourtranstions}, we assumed that there was a dominating color transition that caused the bimodal color distribution that is seen in the Kuiper belt today. We applied colors to the simulated Kuiper belt of \citet{nesvorny_neptunes_2016}, based on where the synthetic KBOs originated before Neptune's migration. This simulated a dominant color change in the primordial disk. The OSSOS survey simulator allows us to make comparisons between the synthetic Kuiper belt and the Col-OSSOS observations. By stepping the color transition position out through the pre-Neptune migration disk we investigate possible initial disk layouts produced the Kuiper belt colors observed by Col-OSSOS.
    
    \subsection{Building the Synthetic KBO Population} \label{sec:simpop}
    
        After the dynamical cuts outlined in Section \ref{sec:modeldynamicalcuts}, the \citet{nesvorny_neptunes_2016} synthetic population contained $\sim$3500 simulated KBOs.The OSSOS survey simulator takes synthetic planetesimals on simulated orbits and biases them to what OSSOS would have detected, and so what Col-OSSOS would have selected to observe with \textit{g-}, \textit{r-} and \textit{J-band} observations. Therefore, we use the simulations as a tool to make comparisons between the dynamical model and the Col-OSSOS observations.
        
        In order to match the number of objects observed in Col-OSSOS within E, H and L blocks, we required a sufficiently large synthetic population to be biased by the survey simulator. \citet{nesvorny_neptunes_2016} tracked their synthetic KBOs from their initial positions to their final positions. Therefore, their final semimajor axis, eccentricity and inclination could not be altered without their initial positions losing meaning. \revision{We therefore duplicated the semimajor axis, eccentricity and inclination components from the \citet{nesvorny_neptunes_2016} model in order to increase the number of synthetic objects. In order to calculate how many synthetic objects were needed we generated possible $H_r$ distributions, which we compared with the $H_r$ distribution observed by OSSOS. By scaling this magnitude distribution we could adjust the number of synthetic objects included in our simulation until it best matched the number observed by OSSOS. We then randomly drew our intrinsic synthetic population from a sample of $\sim$4,000,000 duplicated objects from the final population of the model of \citet{nesvorny_neptunes_2016}.
        }
        
\begin{figure}
            \centering
            \includegraphics[width=\textwidth]{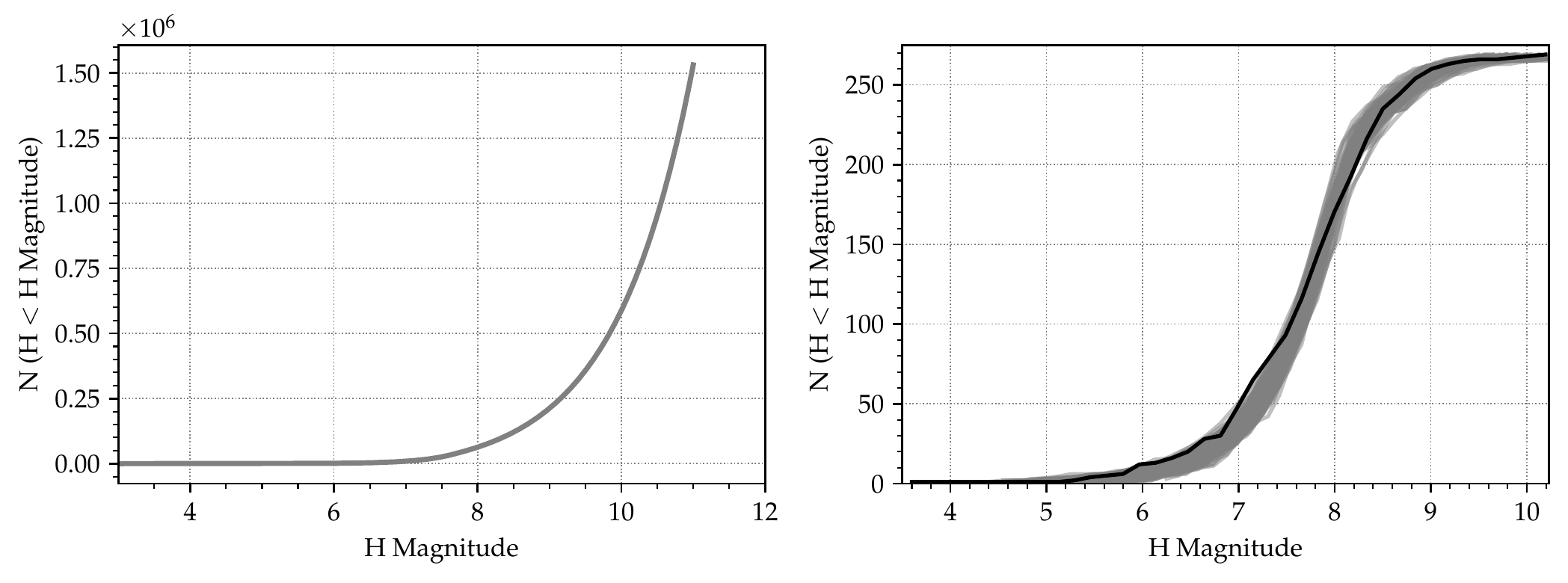}
            \caption{The left plot shows the cumulative $H_r$ magnitude distribution. It follows the broken exponential outlined by Equations \ref{eqn:H_pre} and \ref{eqn:H_post}. The brightest magnitude limit is set by \citet{2008ssbn.book..335B} and the dimmest magnitude limit is set by the limit of the simulated population that can be synthetically detected by the OSSOS survey simulator within the Col-OSSOS magnitude limit. The right plot shows the OSSOS cumulative $H_r$ magnitude distribution in black. In \revision{gray} is the cumulative $H_r$ magnitude distribution of 50 simulated Kuiper belt populations, biased by the OSSOS survey simulator.}
            \label{fig:Hmag}
        \end{figure}

        We generated absolute magnitudes and derived brightnesses using the OSSOS survey simulator. Previous work has shown that the $H_r$ magnitude distribution of hot population KBOs within the Kuiper belt follows a broken exponential with a sharp transition \citep[e.g.][]{fraser_absolute_2014,shankman_ossos_2016,lawler_ossos_h._2018}. The bright end of this $H_r$ magnitude distribution has a more steep slope than the faint end, with the break between these at a $H_r$ magnitude of 7.7 \citep{lawler_ossos_h._2018} as shown in Figure \ref{fig:Hmag}. We use the divot case of the $H_r$ magnitude distribution from \citet{lawler_ossos_h._2018}, with the intensity of the divot given by the contrast $c$. Given that no difference in slope between the color distributions in the hot population has been identified, we made the assumption that the red and neutral surfaced objects followed the same $H_r$ magnitude distribution. \revision{Previous works \citep[e.g.][]{2009Icar..201..284B,2014ApJ...793L...2L,2014ApJ...782..100F} have identified a difference in albedos for red and neutral surface colored KBOs. However, we are working exclusively in $H_r$ magnitude and as stated earlier  the same power law slopes have been observed to describe both the red and neutral colored brightness distributions.} The bright end of the $H_r$ magnitude distribution (with $H_r<7.7$) followed Equation \ref{eqn:H_pre}, with the exponential slopes derived from \citet{lawler_ossos_h._2018}.
        
        \begin{equation}
            \label{eqn:H_pre}
            N(\leq H) = A 10^{\alpha_1 (H_r - H_0)}
        \end{equation}
        
        Where $N(\leq H_r)$ is the cumulative number of objects at $H_r$ magnitude $H_r$, $H_0 = 3.6$ is a normalisation constant with value equal to the brightest $H_r$ magnitude for OSSOS detections, $A$ is a scaling factor equal to the number of objects at $H \leq H_0$, and $\alpha_1 = 0.9$ \citep{fraser_absolute_2014}. The magnitude distribution after the break is described by Equation \ref{eqn:H_post}.
        
        \begin{equation}
            \label{eqn:H_post}
            N(\leq H) = A 10^{\alpha_1 (H_B - H_0)} + B 10^{\alpha_2 (H - H_B)} -B
        \end{equation}
        \begin{equation}
            B = c A \frac{\alpha_1}{\alpha_2} 10^{(\alpha_1(H_B - H_0))}
        \end{equation}
        
        Where $H_B$ is the break in the $H_r$ magnitude distribution \citep[$H_B=7.7$;][]{lawler_ossos_h._2018}, the contrast value $c = 0.85$, and $\alpha_2 = 0.4$ \citep[][]{fraser_absolute_2014}. We chose a $H_r$ magnitude range between 11 and 3. There were no OSSOS targets brighter than an absolute magnitude of $\sim$3; additionally the magnitude distribution below $\sim$3 significantly flattens, and so does not follow the exponentials given in Equations \ref{eqn:H_pre} and \ref{eqn:H_post} \citep{2008ssbn.book..335B}. The limit of 11 was chosen due to there being few OSSOS KBOs (and no Col-OSSOS KBOs) with absolute magnitudes fainter than this value, and the OSSOS survey simulator `detected' few objects dimmer than $\sim$11 mag and so this limit reduced the computing time.
        
        The scaling constant $A$ and contrast $c$ were adjusted until the simulated distribution (after being run through the OSSOS survey simulator) best matched the observed $H_r$ distribution from the non-resonant, non-Centaur, dynamically excited objects in OSSOS.  As shown in Figure \ref{fig:Hmag}, the cumulative numbers of objects with increasing $H_r$ magnitude were compared between the observations and the biased simulations. A Kolmogorov-Smirnov (KS) test was used to identify the best matching $H_r$ distribution. This resulted in values for the scaling factors of $A = 8$ and $B = 75000$. Using these scaling factors we found that $\sim$2,700,000 synthetic KBOs gave us `observed' KBOs with a $H_r$ distribution consistent with that observed by OSSOS.
        
        Once we found the $H_r$ distribution that best matched what OSSOS observed, we assigned these brightnesses to our synthetic KBOs. The semimajor axes, eccentricities and inclinations were taken directly from the \citet{nesvorny_neptunes_2016} particles, after the dynamical cuts described in Section \ref{sec:modeldynamicalcuts}. Values for the longitude of node, the argument of pericenter, and the mean anomaly were also generated for the simulated KBOs. These angles were drawn from a random, uniform distribution between 0${^\circ}$ and 360${^\circ}$. Due to the planetary effects that these bodies have experienced since the end of Neptune's migration it was assumed that their orbital angles have been uniformly \revision{randomized} over $\sim$4 billion years \citep{2006Icar..184...59B}. Surface colors also needed to be allocated to the simulated KBOs. The colors were assigned based on the location in the pre-Neptune migration Kuiper belt those simulated objects originated, an example of which is shown in Figure \ref{fig:nes_colours}. The mean $(g - r)$ and $(r - J)$ magnitudes were taken from the red and neutral colored groups within Col-OSSOS E, H and L blocks, and assigned to our red/neutral synthetic Kuiper belt population. For the neutral surfaced objects the means are $(g - r) = 0.6$ and $(r - J) = 1.4$ and for the red surfaced objects they are $(g - r) = 0.9$ and $(r - J) = 1.6$.
        
    \subsection{Running the Color Simulations}
        
        This simulated Kuiper belt population was input to the OSSOS survey simulator \citet{lawler_ossos:_2018}, and so allowed us to generate a `virtual OSSOS' made up of biased synthetic KBOs within the OSSOS observing fields. From this we selected the synthetic KBOs with discovery magnitude $<$23.6 in the three observing blocks that we were considering (E, H and L blocks), thus creating a `virtual Col-OSSOS' sample of synthetic objects with known colors. The `virtual Col-OSSOS' was then compared with the Col-OSSOS comparison sample outlined in Section \ref{sec:sampleselection}. By creating numerous virtual color surveys for each initial disk layout, we worked out what fraction of these color simulations matched the numbers of each surface color within the Col-OSSOS comparison sample.
        
        We treated each block separately and compared each block individually to the biased synthetic population. When the blocks are looked at together, it is possible that there may be one that is dominating the signal (as each have different characteristics). We want to look individually at each block to ensure that it is not the problem, and then combine them together to reject or accept models. In the comparison sample outlined in Section \ref{sec:sampleselection}, 2013 UM17 in L block has unknown surface colors. We therefore allowed for two scenarios for this object: in scenario A the unknown surface was neutral colored, while in scenario B it was red. In Table \ref{tab:Colour_Numbers} we \revision{summarize} the total number of each surface type in each block, including the alternate numbers for scenarios A and B. Due to the small numbers of each surface color in a given observing block we applied Poisson errors to the numbers of each surface type in each observing block \citep[following the prescription of ][]{1991ApJ...374..344K} and we report a 95\% confidence level as our uncertainty. The uncertainties on the numbers of different surface colors due to these Poisson errors are \revision{summarized} in Table \ref{tab:Colour_Numbers}.
        
        We checked that the biased simulated population matched the possible total number of objects observed by Col-OSSOS in each observing block (within the Poisson uncertainties). This simply ensured that we were only keeping simulated Kuiper belts with total number of KBOs consistent with our Col-OSSOS comparison sample. The total number of KBOs in E, H and L observing blocks is given in Table \ref{tab:Colour_Numbers}, and all three observing blocks had to have total number of KBOs within these limits to be included.  Any simulated Kuiper belt with total number of objects (in any block) outside of these ranges was discounted. For each initial disk layout, we ran color simulations until we had 40,000 cases with this total number of KBOs matching. Of these 40,000 cases that matched the total number of KBOs, we then investigated how many of them also had a total number of each surface color within the Poisson limits on the Col-OSSOS comparison sample (given in Table \ref{tab:Colour_Numbers}). Again, these numbers of each surface color had to be matched in all observing block simultaneously. We used this to work out what percentage of our color simulations matched the Col-OSSOS comparison sample. We repeated this for each position (in steps of 0.5 au) for the color transition in the initial disk, and each color layout (inner neutral / outer red and inner red / outer neutral), allowing us to investigate how the fraction of simulations that matched the Col-OSSOS comparison sample changed with differing initial disks.
        
        \begin{deluxetable}{c|cc|c}
            \tablecaption{\label{tab:Colour_Numbers} Summary of the numbers of red, neutral and unknown colored surfaces among the comparison sample, along with their associated Poisson errors. In scenario A (L\textsubscript{A}) the unknown surface colors are assumed to be neutral colored, and in scenario B (L\textsubscript{B}) they are assumed to be red. In the column `Total' we show the ranges that the total number of KBOs could have in each scenario (where $x$ is the number of KBOs), in each observing block.}
            \tablehead{
                \colhead{Observing Block}  & \colhead{Neutral} & \colhead{Red} & \colhead{Total} }
                \startdata
                E                       & $3^{+5}_{-2}$   & $4^{+5}_{-3}$    & $2\leq x\leq 17$    \\
                H                       & $4^{+5}_{-3}$   & $4^{+5}_{-3}$    & $2\leq x\leq 18$    \\
                L\textsubscript{A}      & $5^{+6}_{-3}$   & $0^{+3}_{-0}$    & $2\leq x\leq 14$    \\
                L\textsubscript{B}      & $4^{+5}_{-3}$   & $1^{+4}_{-1}$    & $1\leq x\leq 14$    \\
              \enddata
        \end{deluxetable}

\section{Results and Discussion} \label{sec:results}
    
\begin{figure}
        \centering
        \includegraphics[width=\textwidth]{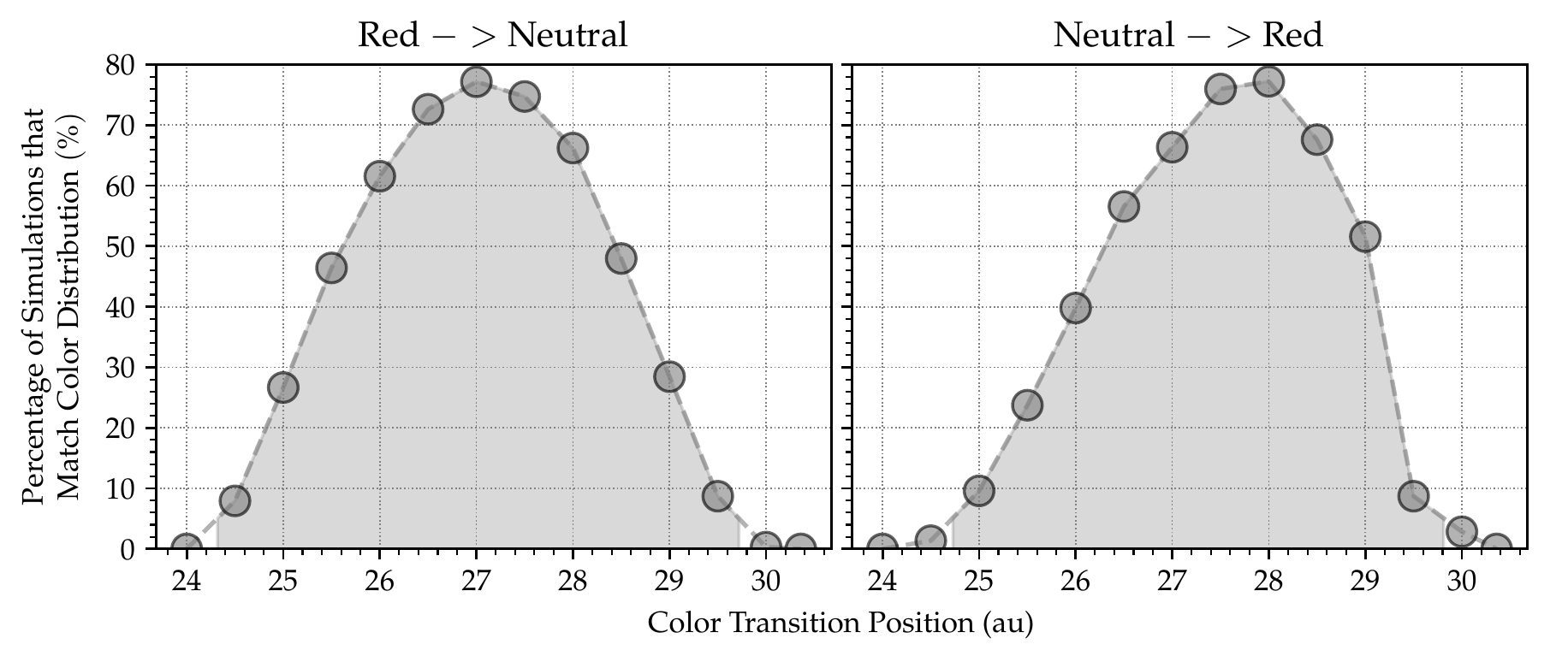}
        \caption{Shows how the percentage of simulations matching the color distribution changes with the position of the synthetic ice line in the primordial planetesimal disk. The initial disk ranges from 24 au to 30.36 au. These probabilities are the the percentage of the 40,000 simulations for each initial disk transition that match the observed surface colors seen in Col-OSSOS. The left plot shows the results for the inner red, outer neutral primordial disk, and the right plot shows for the inner neutral, outer red primordial disk. The \revision{gray} shaded region shows the region that the simulations cannot be ruled out to a $95\%$ confidence level.}
        \label{fig:probabilities}
    \end{figure}

    In Figure \ref{fig:probabilities} we show the fraction of color simulations that are consistent with the Col-OSSOS observations as a function of color transition distance, ranging from 24 au to $\sim$30 au for both potential color layouts (inner neutral / outer red and inner red / outer neutral). We tallied the number of simulations that had red and neutral `detections' consistent with the three Col-OSSOS blocks and plot in Figure \ref{fig:probabilities} the results from each color transition step. We found that the peak percentage of simulations with `detections' consistent with Col-OSSOS E, H and L block numbers is significantly greater than 5\% for at least one of the transition distances in both of the scenarios (inner red / outer neutral and inner neutral / outer red). Therefore, we find both pre-Neptune migration color distributions to be viable.
    
    Based on our results we found the transition distance to be $27^{+3}_{-3}$ au for the inner red / outer neutral disk, using the peak of our measured distribution as the best fit value and the 95\% confidence limits chosen as the transition distances where only 5\% of the simulations matched Col-OSSOS. Similarly, for the inner neutral / outer red disk we found the transition distance to be at $28^{+2}_{-3}$ au. 
    
    As mentioned in Section \ref{sec:colourtranstions} `blue binaries' are proposed to have formed at $\sim$38 au \citep{fraser_all_2017}, separately from both the cold classical KBOs and the dynamically hot KBOs. Assuming that this formation distance is the only way that these `blue binaries' can be produced, and that their neutral surfaces are similar to those in the hot population, our inner red / outer neutral is consistent with the production of `blue binaries'. Although the color transition position for our inner neutral / outer red is not consistent with the origin scenario of the `blue binaries' \revision{within \citet{fraser_all_2017}}, in this work we can only test transitions up to $\sim$30 au. \revision{However,} as our error bar on this disk layout reaches all the way to 30 au, this suggests that transition positions beyond 30 au could be possible \revision{in the potential case that the initial disk extended beyond 30 au.}
    
    \citet{2020AJ....160...46N} have also explored possible color transition positions in the primordial Kuiper belt. They used a model of Kuiper belt formation with a similar primordial disk between 24 au and 30 au, but with the addition of a low mass disk extension from 30 au to 40 au. They only investigated an inner neutral / outer red primordial disk due to the predominantly neutral colored surfaces of Neptune's trojans, and proposed a color transition position between 30 au and 40 au. Similarly, \citet{2021AJ....162...19A} used N-body simulations to find an inner neutral / outer red disk with the color transition between $\sim$38 and 42 au based on the lack of red surface KBOs at higher eccentricities. Both of these works find that the inner neutral / outer red disk are consistent with the modern day Kuiper belt colors, in agreement with our findings. However, as alternate disk models were used, a direct comparison in possible color transition position is difficult. 
    
    Due to the fact that neither the inner red / outer neutral nor the inner neutral / outer red disk layouts could be ruled out at this point, we investigated the color/inclination distribution. \citet{2019AJ....157...94M} showed that color and inclination in the Kuiper belt are correlated, with higher inclined KBOs tending to have more neutral surface colors. Therefore we performed a two dimensional Kolmogorov Smirnov test on the color and inclination, so as to calculate the largest absolute difference between our simulations and the Col-OSSOS comparison sample. As we assigned discrete red and neutral $(g-r)$ values to our simulated KBOs the colors were either a $(g-r)$ value of 0.9 for red KBOs, or 0.6 for neutral colored KBOs. The inclinations assigned were simply the inclinations from the dynamical model for the simulated KBOs, and the observed inclinations for the Col-OSSOS comparison sample.
    
    We performed 2D KS tests for the most likely color transition position in each initial disk layout; 27 au for the inner red / outer neutral disk and 28 au for inner neutral / outer red primordial disk. For each layout we generated 5000 simulated populations that matched the numbers of the Col-OSSOS color distribution when biased by the OSSOS survey simulator. We then created a `supersample' of the orbital inclinations and surface \revision{colors} for each disk layout. In order to calibrate our 2D KS test results, we initially tested the Col-OSSOS comparison sample against the full supersample. From our supersample we then drew subsamples of the same size as our Col-OSSOS comparison sample, and performed 2D KS tests between these subsamples and the supersample. 
    
    These 2D KS tests provided a statistical measure of the maximum difference between the color simulations and the Col-OSSOS comparison sample in color/inclination space. For the inner red / outer neutral primordial disk we found that 8.45\% of the subsample versus super sample tests had D statistic less than the comparison sample versus supersample test, therefore we could not reject the hypothesis that they were drawn from the same distribution. In the case of the inner neutral / outer red primordial disk only 5\% of the subsample versus super sample 2D KS tests had D statistic less than the comparison sample versus supersample test, and therefore we could also not reject the hypothesis that they were drawn from the same distribution in this scenario. Therefore, based on the color and inclination distributions neither of the initial disk layouts can be ruled out. 
    
\section{Conclusions}
    
    In this work we used a dynamical model of the Kuiper belt's formation through Neptune's migration \citep{nesvorny_neptunes_2016} to investigate the location of a dominant color changing ice line in the primordial Kuiper belt. We compared these color simulations with photometry from Col-OSSOS \citep{schwamb_col-ossos:_2019}. We investigated both an inner red / outer neutral initial disk \citep[due to the prescence of `blue binaries' within the Kuiper belt,][]{fraser_all_2017} and an inner neutral / outer red disk \citep[due to the predominantly neutral surface colors of Neptune's Trojans,][]{2013AJ....145...96P,2018AJ....155...56J}. Assuming that the distribution from \citet{nesvorny_neptunes_2016} was accurate, we find that both inner red / outer neutral and the inner neutral / outer red configurations were consistent with the results from Col-OSSS E, H and L blocks. For the inner neutral / outer red primordial disk the ice line is at $28^{+2}_{-3}$ au, to a 95\% confidence level. For an inner red / outer neutral primordial disk the ice line is located at $27^{+3}_{-3}$ au, to a 95\% confidence level. A 2-dimensional KS test was used to investigate the correlation between inclination and surface color in the simulations, and confirmed that neither initial disk layout can be ruled out based on these color simulations. This strongly implies that, for this case, Neptune is efficiently scattering objects throughout the Kuiper belt irrespective of distance.
    
    The differing albedos of different KBO surfaces presents a potential limitation to this work. As discussed in Section \ref{sec:simpop}, we assume that the red and neutral surfaced KBOs follow the same $H_r$ magnitude distribution. Previously, it has been show that the red and neutral colored KBOs have differing surface albedos \citep[e.g.][]{2009Icar..201..284B,2014ApJ...793L...2L,2014ApJ...782..100F}. This could potentially influence the the conversion between our absolute $H_r$ magnitudes and the sizes of the KBOs.
    
    \revision{Our results are dependent on our chosen simulated primordial disk accurately reflecting the state and evolution of the planetary disk and Neptune's migration history. Our chosen dynamical model has a truncated initial disk at 30 au, and our analyzed sample of Col-OSSOS colors  was unable to rule out this scenario. Recent works have explored dynamical models with a low mass disk extension beyond 30 au \citep[e.g.][]{2020AJ....160...46N,2021AJ....162...19A}). In the case of the inner  neutral / outer red disk, the most probable color transition may have been pushed out further than 28 au if we had used a dynamical model with an extended disk. However, even in the case that our  chosen dynamical model does not reflect our Kuiper Belt’s formation history we have still placed a lower limit on the possible color transition positions that produce the modern day Kuiper Belt colors. Additionally, \citet{2021A&A...650A.161P} show that the early inward migration of an accreting Neptune could emplace ‘blue binaries’ into the cold classical region before Neptune’s planetesimal driven migration phase. Therefore, this provides a potential additional avenue through which the ‘blue binary’ KBOs can be produced. Consequently, we cannot rule out either of our initial disk layouts based on their ability to produce the ‘blue binary’ KBOs. \citet{2015AJ....150...73N} suggest that the low inclination hot classical KBOs (those with inclinations below 12$^\circ$) originated from a disk extended beyond 30 au. As we are testing a dynamical model with no significant material beyond 30 au contributing to the modern day  Kuiper Belt, the lack of correlation between color and inclination may have been due to the range of  our initial disk. In the future, it may be useful to combine the full Col-OSSOS sample with potential future color surveys and dynamical models to further investigate possible color transition positions in these different scenarios.}
    
\acknowledgments

    LEB acknowledges funding from the Science Technology Funding Council (STFC) Grant Code ST/T506369/1. MTB appreciates support during OSSOS from UK STFC grant ST/L000709/1, the National Research Council of Canada, and the National Science and Engineering Research Council of Canada. KV acknowledges support from NASA (grants NNX15AH59G and 80NSSC19K0785) and NSF (grant AST-1824869). \revision{MES acknowledges support from STFC grant ST/V000691/1. This work was supported by the Programme National de Plantologie (PNP) of CNRS-INSU co-funded by CNES.} The authors acknowledge the sacred nature of Maunakea and appreciate the opportunity to obtain observations from the mountain.  The observations were obtained as part of observations from the programs (GN-2014B-LP-1, GN-2015A-LP-1, GN-2015B-LP-1, GN-2016A-LP-1, GN-2017A-LP-1, GN-2018A-Q-118, GN-2018A-DD-104, GN-2020B-Q-127, and GN-2020B-Q-229) at Gemini Observatory. The international Gemini Observatory is a program of NSF’s OIR Lab, and is managed by the Association of Universities for Research in Astronomy (AURA) under a cooperative agreement with the National Science Foundation. On behalf of the Gemini Observatory partnership: the National Science Foundation (United States), National Research Council (Canada), Agencia Nacional de Investigación y Desarrollo (Chile), Ministerio de Ciencia, Tecnología e Innovación (Argentina), Ministério da Ciência, Tecnologia, Inovações e Comunicações (Brazil), and Korea Astronomy and Space Science Institute (Republic of Korea). The GMOS-N observations were acquired through the Gemini Observatory Archive at NSF’s NOIRLab and processed using DRAGONS (Data Reduction for Astronomy from Gemini Observatory North and South). We are grateful for use of the computing resources from the Northern Ireland High Performance Computing (NI-HPC) service funded by EPSRC (EP/T022175).

\bibliography{ColourPaper}{}
\bibliographystyle{aasjournal}

\end{document}